# QUBIT DATA STRUCTURES FOR ANALYZING COMPUTING SYSTEMS

Vladimir Hahanov[1], Wajeb Gharibi[2], Svetlana Chumachenko[1] and
Eugenia Litvinova[1]

[1]Department of Computer Engineering, Kharkov National University of
Radioelectronics, Kharkov, Ukraine
hahanov@kture.kharkov.ua
[2] Jazan University, Kingdom of Saudi Arabia,
gharibiw2002@yahoo.com

## ABSTRACT

*Qubit models and methods for improving the performance of software and hardware for analyzing digital devices through increasing the dimension of the data structures and memory are proposed. The basic concepts, terminology and definitions necessary for the implementation of quantum computing when analyzing virtual computers are introduced. The investigation results concerning design and modeling computer systems in a cyberspace based on the use of two-component structure <memory - transactions> are presented.*

## 1. INTRODUCTION

Market feasibility of emulation of quantum computing methods to create virtual (cloud) computers (VC) in a cyberspace is based on the use of qubit data models, focused on parallel solving discrete optimization problems through significant increase in memory costs. We do not consider the physical basis of quantum computing, originally described in the works of scientists, focused on the use of non-deterministic quantum interactions within the atom. We do not address the physical foundations of quantum mechanics, concerning non-deterministic interactions of atomic particles [1-4], but we use the concept of qubit as a binary or multivalued vector for a concurrent definition of the power set (the set of all subsets) of the states for the discrete cyberspace area based on linear superposition of unitary codes, focused on parallel executing methods for analyzing and synthesizing cyberspace components.

In the market of electronic technologies there is competition between basics of idea implementation [2]: 1) Soft implementation of the project related to the synthesis of interpretative model of the device or hardware implementation of programmable logic devices based on FPGA, CPLD. There are advantages in manufacturability design modifications, the disadvantages - low performance in the operation of a digital system. The advantages lie in the manufacturability of design modifications, the disadvantages – in low performance of a digital system. 2) Hard implementation is focused to the use of compilation models in developing software applications or in the implementation of the project on the basis of VLSI chips. The advantages and disadvantages of hard implementation are inverted with respect to soft projects: high performance and impossibility of modification. Given above mentioned four basic variants for the implementation of the idea below are represented quantum data structures, focused on raising the performance of flexible models of software or hardware project implementation.

## 2. QUANTUM STRUCTURES FOR DESCRIBING DIGITAL SYSTEMS

n-Qubit is a vector form of unitary encoding the universe of n primitives to specify the power set of states $2^{2^n}$ by using $2^n$ binary variables.

For example, if n=2, then the 2-qubit sets 16 states by using 4 variables. If n=1, the qubit sets 4 states on the universe of two primitives by using 2 binary variables ( 00,01,10,11 ) [5] . Herewith, the superposition (simultaneous existence) of $2^n$ states in a vector is supposed. Qubit (n-qubit) allows using the logical operations instead of set-theoretic ones to significantly speed up the analysis of discrete systems. Further the qubit is identified by the n-qubit or vector if this does not prevent the understanding of presented material. As quantum computing is related to analysis of qubit data structures, further we use definition "quantum" for identifying technologies, based on two properties of quantum mechanics: concurrency of processing and superposition of states. When defining logical functionality the following synonym of qubit is used: Q-coverage (Q-vector) [5] is used as a unified vector form for the definition of superposed output states corresponding to unitary codes of addresses for input variables of any logical function.

Qubit of a digital system is a form for defining a structural primitive that is invariant to the technologies for implementing the functionality (hardware, software). Moreover, "quantum" synthesis of digital systems based on qubit structures is not rigidly tied to the Post's theorem, defining the conditions for the existence of a functionally complete basis. At the proposed abstraction level n-qubit gives exhaustive and wider opportunities for vector defining any function of the set $\beta(f) = 2^n$. Format of the structural qubit component of the digital circuit $Q^* = (X, Q, Y)$ involves interface (input and output variables), as well as qubit-vector Q, defining the functionality $Y = Q(X)$, the dimension of which is defined by the power function of the number of input lines $k = 2^n$.

Practically oriented novelty of qubit modeling lies in replacing truth tables of digital device components by vectors of output states. Such transformations can be simply demonstrated for the logic element. Let functional primitive has the following binary coverage:

$$P = \begin{array}{|cc|c|} \hline X_1 & X_2 & Y \\ \hline 0 & 0 & 1 \\ 0 & 1 & 1 \\ 1 & 0 & 1 \\ 1 & 1 & 0 \\ \hline \end{array} ,$$

which can be transformed by unitary encoding input vectors based on the use of two-stroke alphabet [2, 5]. It was originally designed for a compact description of all possible transitions of automaton variables, as illustrated in Fig. 1 by corresponding graph and interpretation of symbols.

Here symbols, their binary and unitary codes are presented (for instance, $Q = 00 - 1000$) for describing the two adjacent states of automatic variables. Structurally alphabet is the set of all subsets of the states on the universe of four primitives Y={Q,E,H,J}. A unitary code corresponds to the format of the vector comprising two qubits by using which 16 symbols of two-stroke alphabet are generated. By using two-stroke alphabet any coverage of functional two-input logic primitive can be represented by two or even one cubes, given that they are mutually inverse:

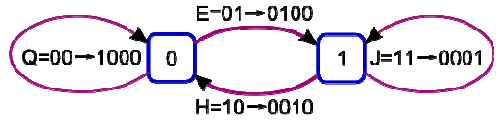

$$P = \begin{array}{|c|c|} \hline 00 & 1 \\ \hline 01 & 1 \\ \hline 10 & 1 \\ \hline 11 & 0 \\ \hline \end{array} = \begin{array}{|c|c|} \hline Q & 1 \\ \hline E & 1 \\ \hline H & 1 \\ \hline J & 0 \\ \hline \end{array} = \begin{array}{|c|c|} \hline V & 1 \\ \hline J & 0 \\ \hline \end{array} = \begin{array}{|c|c|} \hline 1110 & 1 \\ \hline 0001 & 0 \\ \hline \end{array} \rightarrow \begin{array}{|c|c|c|c|} \hline 1 & 1 & 1 & 0 \\ \hline \end{array}$$

| Q = 00 → 1000 | E = 01 → 0100 | H = 10 → 0010 | J = 11 → 0001 |
|---|---|---|---|
| O = {Q, H} = <br> = {00,10} → 1010 | I = {E, J} = <br> = {01,11} → 0101 | A = {Q, E} = <br> = {00,01} → 1100 | B = {H, J} = <br> = {01,11} → 0101 |
| S = {Q, J} = <br> = {00,11} → 1001 | P = {E, H} = <br> = {01,10} → 0110 | C = {E, H, J} = <br> = {01,10,11} → 0111 | F = {Q, H, J} = <br> = {00,10,11} → 1011 |
| L = {Q, E, J} = <br> = {00,01,11} → 1101 | V = {Q, E, H} = <br> = {00,01,10} → 1110 | Y = {Q, E, H, J} = <br> = {00,01,10,11} → 1111 | ∅ = 00 → 0000 |

Figure 1. Two-stroke alphabet of automaton variables

At first all the pairs are encoded by the symbols of two-stroke alphabet and then the union of the first three cubes is made according with the rule "co-edge" operator [3]: vectors differing in one coordinate are minimized in one vector. Further the resulting coverage of two cubes is encoded by qubit vectors, corresponding to the given symbols. For modeling fault-free behavior it is enough to have only one cube (zero or unit one), since the second one is always a complement to the first. Consequently, for example, if we choose the unit cube determining 1 in the output, we can remove the bit of output primitive status, reducing the dimension of the cube or primitive model up to the number of addressable primitive states of the element, where address is the vector composed of the binary values of the input variables, by which output state of the primitive is determined. By analogy any truth table can be led to the qubit functionality in the form of vector of output states of logic element of n inputs.

The modeling procedure for Q-vector of the functionality is reduced to writing the bit status in the output variable Y; the bit address is formed by concatenation of values of input variables: $Y = Q(X) = Q(X_1 * X_2 ... * X_j ... * X_k)$. For modeling digital systems components of which are Q-primitives interrelated on the basis of M-vector of equipotential lines, the processing procedure is defined by the equation: $M(Y) = Q[M(X)] = Q[M(X_1 * X_2 ... * X_j ... * X_k)]$. Taken into account numbering of Q-primitives the universal procedure for modeling the current i-element has the following format: $M(Y_i) = Q_i[M(X_i)] = Q_i[M(X_{i1} * X_{i2} ... * X_{ij} ... * X_{ik_i})]$. In this case, the algorithm for analyzing digital system is greatly simplified and the performance of interpretative modeling is improved in $2^n$ times by increasing the amount of memory to describe the functionality of a circuit structure.

Synthesis of Q-coverage for a digital system is reduced to performing superposition of Q-vectors of the functionalities included in it. For example, for three primitives (and, and-not, and-not), which compose a circuit, the operation of superposition generates a Q-vector of whole functionality, where its dimension is greater than the sum of Q-coverages of the original primitives:

$$\begin{array}{c}a\\b\\d\\e\end{array}\begin{array}{|c|}\hline 0\ 0\ 0\ 1\ c\\\hline 1\ 1\ 1\ 0\ f\\\hline\end{array}\ \begin{array}{|c|}\hline c\\f\end{array}\begin{array}{|c|}\hline 1\ 1\ 1\ 0\ g\\\hline\end{array}=\begin{array}{c}a\\b\\d\\e\end{array}\begin{array}{|c|}\hline 1\ 1\ 0\ 1\ 1\ 1\ 0\ 1\ 1\ 1\ 0\ 0\ 0\ 0\ 1\ g\\\hline\end{array}\ .$$

But at that, modeling procedure for a Q-vector of the structure has a higher performance, because it supposes only one access to the Q-coverage for reading the content of a single cell instead of three ones, when a system is represented by three primitives.

Three-element circuit, represented by Q-vectors above, can be defined by the circuit (useful for human), where corresponding decimal numbers are used instead of the vectors:

$$\begin{array}{c}a\\b\\d\\e\end{array}\begin{array}{|c|}\hline 1\ c\\\hline 14\ f\\\hline\end{array}\ \begin{array}{|c|}\hline c\\f\end{array}\begin{array}{|c|}\hline 14\ g\\\hline\end{array}=\begin{array}{c}a\\b\\d\\e\end{array}\begin{array}{|c|}\hline 34679\ g\\\hline\end{array}\ .$$

When processing such form of functional coverage it is necessary to transform a binary code to a decimal vector and calculate the cell address, the contents of which determines the state of the output variable, in this case – g. Naturally, the decimal code exists on paper and in the computer this representation is always a binary vector. In fact, the "soft" circuitry for identifying (numbering) interconnects is perspective, because it is not associated with connecting wires, which are replaced by addresses or line numbers, creating flexible structure of a digital product for replacing primitives in case of detection of design errors or defects.

Qubit representation of functional elements also allows introducing new circuit designations associated with the decimal number of Q-vector defining the functionality. If the system of logical elements includes n=2 inputs, then the number of all possible functions is equal to $k = 2^{2^n}$, where the types or numbers of functionals are presented in the bottom row of the following table:

| 00 | 0 | 0 | 0 | 0 | 0 | 0 | 0 | 0 | 1 | 1 | 1 | 1 | 1 | 1 | 1 | 1 |
|---|---|---|---|---|---|---|---|---|---|---|---|---|---|---|---|---|
| 01 | 0 | 0 | 0 | 0 | 1 | 1 | 1 | 1 | 0 | 0 | 0 | 0 | 1 | 1 | 1 | 1 |
| 10 | 0 | 0 | 1 | 1 | 0 | 0 | 1 | 1 | 0 | 0 | 1 | 1 | 0 | 0 | 1 | 1 |
| 11 | 0 | 1 | 0 | 1 | 0 | 1 | 0 | 1 | 0 | 1 | 0 | 1 | 0 | 1 | 0 | 1 |
| f = | 0 | 1 | 2 | 3 | 4 | 5 | 6 | 7 | 8 | 9 | 10 | 11 | 12 | 13 | 14 | 15 |

Moreover, on the basis of a set of qubits of the first level, defining functions of two variables, we can enter a qubit for unitary encoding two-input functions that allows creating a structure for simultaneous defining and analyzing all the states of the discrete system, where the input variables are the functionals of the first level:

| 00 | 0 | 0 | 0 | 0 | 0 | 0 | 0 | 0 | 1 | 1 | 1 | 1 | 1 | 1 | 1 | 1 |
|---|---|---|---|---|---|---|---|---|---|---|---|---|---|---|---|---|
| 01 | 0 | 0 | 0 | 0 | 1 | 1 | 1 | 1 | 0 | 0 | 0 | 0 | 1 | 1 | 1 | 1 |
| 10 | 0 | 0 | 1 | 1 | 0 | 0 | 1 | 1 | 0 | 0 | 1 | 1 | 0 | 0 | 1 | 1 |
| 11 | 0 | 1 | 0 | 1 | 0 | 1 | 0 | 1 | 0 | 1 | 0 | 1 | 0 | 1 | 0 | 1 |
| Q = | 0 | 1 | 1 | 1 | 1 | 1 | 0 | 0 | 0 | 0 | 0 | 0 | 1 | 1 | 0 | 0 |

This table summarizes four vector-primitives of input variables (00,01,10,11), forming $k = 2^{2^2} = 2^4$ a complete set of all possible functions, which are considered as second-level primitives. Then primitive vectors of the output variables are represented (16 columns from 0000 to 1111); they form $k = 2^{2^4} = 2^{16}$ functional primitives, which are parts of a more complex discrete system and can be analyzed in parallel! Further it is possible to extrapolate creation of more complex system of qubits, where the vector Q=0111110000001100 represented by the bottom row will be considered as one of $k = 2^{2^{16}}$ primitives of third hierarchy level. In each hierarchy level of qubits the number of states (the power set) is exponentially dependent on the number of primitives-vectors $k = 2^{2^{n}}$. If the vector Q includes all unit values Q=1111111111111111, it simultaneously determines the space containing 16 symbols of two-stroke alphabet, which correspond to the power set on the universe of four primitives [3,5].

The main innovative idea of quantum computation compared to the von Neumann machine is to move from the computational procedures of the byte operand, defining a single solution (point) in the discrete space to the quantum parallel processes of qubit operand, at the same time forming the Boolean of solutions. In this thesis the future of high-performance computers for parallel non-digit analysis and synthesis of structures and services for discrete cyberspace are formulated. In other words, the computational complexity of the procedure for processing a set of n elements in the "quantum" processor and a single element in von Neumann machine are equal due to respective n-fold increase in the hardware complexity of the quantum structure.

## 3. GRAPH STRUCTURES FOR DESCRIBING DIGITAL CIRCUITS

A somewhat different circuitry, not based directly on the transistors can be presented in the form of graph structures, where each node (arc) is identified with the functional transformation, which is given by a Q-vector. The arc (node) defines the relationship between the functional Q-coverages, as well as input and output variables. The implementation of such structures is based on the memory cells (LUT FPGA), which are capable to store information in the form of Q-vector, where each bit or digit has an unique address, identified with the input word. However, the software implementation of the structures is competitive in the EDA market in speed due to address implementation of modeling processes for the functional primitives. In addition, hardware support for EDA systems in the form of Hardware Embedded Simulator (HES, Aldec) [5] acquires a new motivation for system-level design of digital products, when software and hardware solutions have the same qubit format. Below (Fig. 2) a combinational circuit, containing 6 primitives and three different logic elements is proposed for consideration.

Three generic graph forms of digital functionality, which use Q-vectors to define the behavior of logical primitives, correspond to the circuit and are shown in Fig. 3. The structure shown in Fig. 3a contains 12 lines (arcs) associated with the quantum functionalities (1=0001 , 7=0111, 14=1110). It is similar to the traditional structural-functional model of a combinational circuit. The graph in Fig. 3b like Sharshunov's register transfer model [2], which is reversed first structure. Here, blue horizontal arcs are identified with functionalities, and nodes – with the groups of input lines for functionalities, combined in register variables by green vertical arcs. States of these lines form a binary vector used as an address for calculating the state of logic element or more complicated functional. The variables used in forming an address for a Q-vector of the functionality can be combined into a single node, with showing all the identificators of lines, which form vector-address. The register graph of the combinational circuit is ranked by levels of formation of input signals, enabling conditions for concurrent handling of the elements of the same level and performability of Seidel iterations [2], which improve the performance of algorithms for fault-free simulating digital systems.

The structure, represented in Fig 3b, is interesting by its register implementation that can be used to formalize the descriptions of both software and hardware models of gate, registry and system levels. This presentation is difficult for perceiving by a person, but it is technological and easily "understood" by computer to automatically create software systems for analyzing and synthesizing computing structures and cyberspace services. Thus, the quantum-register graph of a digital circuit is discontinuous (by galvanic connections) flexible system of interconnected addressed primitives for creating the functional structure of any complexity, especially on PLD, where all combinational primitives are implemented by memory elements (LUTs), which provides high operational performance and online repairing the logic modules.

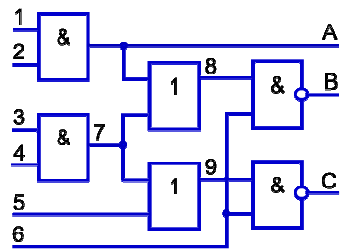

Figure 2. Combination structure of logical primitives

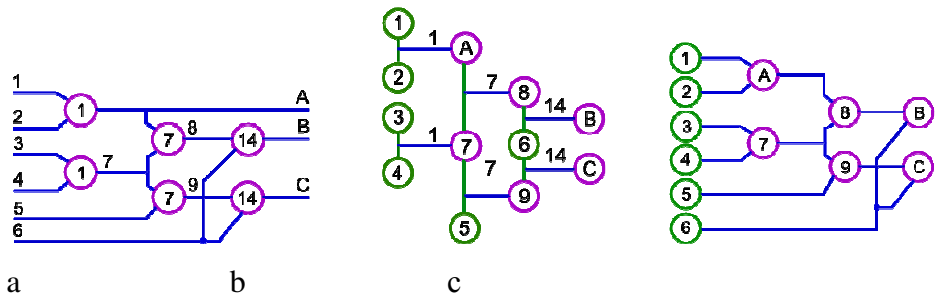

Figure 3. Graph forms of quantum functionalities

A one-dimensional Q-vector is a vector describing the functionality. It can be associated with the output (internal) line of the unit, which is formed in the process of simulation of considered Q-coverage. The register implementation of a combination unit can be represented by the modeling vector M. The functionality with arcs is associated with non-input lines going from the input variables, the values of which form an address of Q-vector bit that forms the state of non-input line under study (Fig. 3c). Otherwise, if the functionality is described by single output primitives, each of them can be identified by number or coordinate of non-input line associated with the element. If the functionality is multi-output, Q-coverage is represented by a matrix with the number of rows equal to the number of outputs. The preference of such primitive lies in the parallelism of concurrent computing the states of several outputs in one access to the matrix at the current address! This fact is an important argument in favor of synthesis of generalized qubits for fragments of a digital unit or whole circuit for their parallel processing in a single time frame.

Closed to the ideal data structure in terms of compactness and processing time, where Q-vectors of functionalities and numbers of input variables are associated with non-input lines of the unit, is the following table:

| L | 1 | 2 | 3 | 4 | 5 | 6 | 7 | 8 | 9 | A | B | C |
|---|---|---|---|---|---|---|---|---|---|---|---|---|
| M | 1 | 1 | 1 | 1 | 1 | 0 | 0 | 1 | 1 | 1 | 0 | 1 |
| X | . | . | . | . | . | . | 34 | A7 | 75 | 12 | 86 | 96 |
| Q | . | . | . | . | . | . | 0 | 0 | 0 | 0 | 1 | 1 |
|  | . | . | . | . | . | . | 0 | 0 | 1 | 1 | 1 | 1 |
|  | . | . | . | . | . | . | 0 | 0 | 1 | 1 | 1 | 1 |
|  | . | . | . | . | . | . | 1 | 1 | 1 | 1 | 0 | 0 |

It shows external variables of a digital circuit, how many functional primitives are available in the structure, and which inputs are associated with each Q-vector. The advantage of the table is the absence of the vector of output numbers for each primitive, but it is still a need to have numbers of input variables for generating addresses, processing of which is time-consuming. Model for analyzing the circuit structure is simplified to calculating two addresses (!) when forming the modeling vector $M_i = Q_i[M(X_i)]$ by eliminating the complex address of the primitive output in writing output states to the coordinates of the vector M.

The qubit-register graph of Fig. 3c can be represented as matrix $\mu = \left| \mu_{ij} \right|, \ i = \overline{1,p}; \ j = \overline{1,q}$ for parallel-to-serial processing logic primitives:

| $\mu_{ij}$ | 1 | 2 | 3 |
|---|---|---|---|
| 1 | $\boxed{\frac{1}{2}}$ $\boxed{1}$ $\boxed{A}$ | $\boxed{\frac{A}{7}}$ $\boxed{7}$ $\boxed{8}$ | $\boxed{\frac{8}{6}}$ $\boxed{14}$ $\boxed{B}$ |
| 2 | $\boxed{\frac{3}{4}}$ $\boxed{1}$ $\boxed{7}$ | $\boxed{\frac{7}{5}}$ $\boxed{7}$ $\boxed{9}$ | $\boxed{\frac{6}{9}}$ $\boxed{14}$ $\boxed{C}$ |
| 3 | $\boxed{\frac{X}{X}}$ $\boxed{1}$ $\boxed{X}$ | $\boxed{\frac{X}{X}}$ $\boxed{7}$ $\boxed{X}$ | $\boxed{\frac{X}{X}}$ $\boxed{14}$ $\boxed{X}$ |

which shows the interaction of Q-coverages at three operation levels in accordance with the format (X–Q–Y) inputs-Q-vector-output for each primitive: [(1,2–1–A), (3,4–1–7)], [(A,7–7–8), (7,5–7–9)], [(8,6–14–B), (6,9–14–C)]. To provide the correctness of the functionality, it is necessary to generate all input variables till a given moment. Therefore quantum-register graph is split into operation levels, where all primitives within a single level can be processed in parallel, and the levels – in succession. Qubit matrix due to its regular structure is focused on solving the following problems: 1) Repair of logical primitives in the operation due to readdressing faulty elements on spare primitives (line 3) [2], just as is done in the memory matrix; 2) Index addressing each quantum of the matrix $\mu_{ij} \in \mu, \mu_{ij} = (X_{ij}, Q_{ij}, Y_{ij})$ for rapid repair of failed primitives (in the example we can replace three faulty primitives, one of each layer); 3) Providing high performance of combinational unit prototype based on quantum primitives implemented in PLD LUTs [2] due to parallel processing primitives of a single layer; 4) Developing a matrix quantum multi-processor, focused on synthesis of hardware prototypes of combinational units of large dimension to significantly speed up testing and verification of digital systems on chips like Aldec Hardware Embedded Simulator (HES) [2, 5]; 5) Developing methods of analysis and synthesis of combinational circuits, focused to matrix realizing quantum structures of logic elements by means of their implementation in PLD memory elements; 6) Developing a code generator for implementing the quantum matrix of the combinational circuit in the structure of PLD circuit primitives; 7) Designing a control automaton for functional processing and repairing the quantum matrix of combinational unit implemented in PLD structure.

The model of control automaton for simulating qubit structure of the combinational circuit involves three items:

1. The initiation of the next input action for combinational unit.

2. Selection of the next layer (matrix column) with the number i for parallel processing qubit primitives Q to form output states at the address of an input word represented by the vector $M(X_{ij})$, where $X_{ij}$ is a vector of numbers of input variables for the primitive $Q_{ij}$, M is modeling vector for all lines of combinational unit: $M(Y_{ij}) = Q_{ij}[M(X_{ij})]$, $j = \overline{1, q}$.

3. Incrementing the index column i=i+1 and going to the item 2 for processing the next layer of qubit primitives. After the analysis of all the columns of the matrix $i = p$ incrementing index of the next input pattern t=t+1 is performed, and subsequent going to the item 1. When reaching a finite number of input patterns $t = n_{max}$ the loop for processing test of the qubit matrix ends.

The hardware implementation of the quantum structure of the digital devices, based on the use of memory elements, is shown in Fig. 4. The structure of the circuit contains the following variables and functional elements: input is designed for serial entering input values of vector M; rst – general reset of the system (in this case for counters); clk – sync input; counter of inputs – counter for filling the input coordinates of the vector M; counter of element – counter of processed primitive number, which provides two cycles for reading the input set of two coordinates of the vector M; Q[3:1] – bus for number of processed primitive; Q[0] – variable for mode of reading input value from the vector M or writing the result to M. Memory: Ram 8x4 output – stores the number of primitive output lines; Ram 8x4 input 1 and Ram 8x4 input 2 – store the numbers of primitive input lines. Ram 16x1d – dual-port memory for storing the modeling vector M, where addr0 – address of the input 1 when the value 00 appears on control inputs of the multiplexer, address of writing result when the value 01 appears on control inputs of the multiplexer, address for initializing input data when the value 1X appears on control inputs of the multiplexer; addr1 – address of input 2 for processed primitive; di0 – memory data input when processing primitive (MUX=1) or external input when initializing input data (MUX=0); we – permission of writing in the vector M; do0 – output, corresponding to the input addr0; do1 – output, corresponding to the input addr1. RAM 32x1 is designed for storing Q-vectors defined functionalities of combinational circuit: di – data input that can be used to initialize (write) the structure of quanta; addr – [4:0]: addr[4:2] – element number, addr[1:0] – input set for a primitive.

The complexity of hardware implementation of the combinational circuit is 150 gates, which include 20 LUTs of Xilinx Spartan 3E element system. The speed of the operation or generation of the modeling vector is 180 ns.

## 4. CONCLUSION

The qubit structure of combination devices provides an opportunity to make a simple automaton from the combinational circuit (integrating memory, functionality quanta, transaction operation) and move from the software simulation of digital systems for the hardware emulation of structures and processes, which are invariant with respect to implementation technologies. An analogue and prototype is Dr. Stanley Hyduke's hardware accelerator of simulation processes PRUS (Aldec Inc.) [2, 5] focused on reducing the time of the design and verification of digital systems on chips. But it is proposed to use the processor for soft (address based) hardware simulating qubit structures for the direct functional purpose as a computing product, delivering services to the consumer. This maintains the high speed operation of the device, supplemented by an opportunity of repairing in real time that is important for critical system.

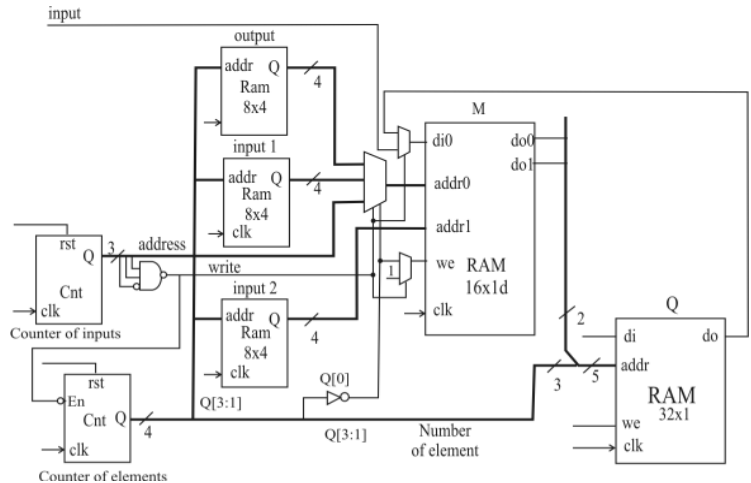

Figure 4. Hardware implementation of qubit structure of combinational circuit

1. Qubit models for describing digital systems and components are proposed, which are characterized by compactness of description of the truth tables in the form of Q-coverages due to the unitary encoding input states, which gives an opportunity to improve the performance of software and hardware applications for the interpretative simulation of computing devices due to address analysis of logic primitives.

2. The matrix model of qubit primitives for implementing combinatorial circuits is represented; it is characterized by address combining of Q-coverages by using memory elements, which are softly connected in the digital circuit through the state vector of lines that enables repairing faulty logic primitives in real-time by way of re-addressing them on spare components at sufficiently high speed operation of the computing device.

3. An innovative idea of quantum computation is described, which is characterized by the transition from the computational procedures with byte operand, defining a single solution (point) in a discrete space, to the logical register parallel processes with the qubit operand, simultaneously generating a power set of solutions that enables to define new perspectives of creating high-performance computers for parallel analyzing and synthesizing discrete structures and services in cyberspace.

## REFERENCES

[1] Michael A. Nielsen & Isaac L. Chuang, (2010) *Quantum Computation and Quantum Information*, Cambridge University Press.

[2] Hahanov V.I., (2009) *Digital System-on-Chip Design and Test*, Kharkov: Novoye Slovo.

[3] Hahanov V.I., (1995) *Technical diagnosis of digital and microprocessor structures*, K.: ICIO.

[4] Gorbatov V.A. (1986) *Fundamentals of Discrete Mathematics*, M.: Vysshaya Shkola.

[5] Hahanov V.I., Murad Ali Abbas, Litvinova E.I., Guz O.A., Hahanova I.V., (2011 ) "Quantum models of computing processes", *Radioelectronics & Informatics*, No. 3, pp35-40.

**Authors**

Vladimir Hahanov – Professor, Doctor of Science, IEEE Senior Member, IEEE Computer Society Golden Core Member, Dean of Computer Engineering Faculty, Kharkov National University of Radioelectronics, Ukraine.

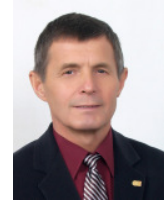

Wajeb Gharibi – PhD, Professor of Jazan University, Kingdom of Saudi Arabia.

Svetlana Chumachenko – Professor, Doctor of Science, IEEE Member, Head of Department "Computer Aided Design of Computers", Kharkov National University of Radioelectronics, Ukraine.

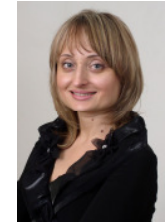

Eugenia Litvinova – Professor, Doctor of Science, IEEE Member, Vice Dean of Computer Engineering Faculty, Kharkov National University of Radioelectronics, Ukraine.

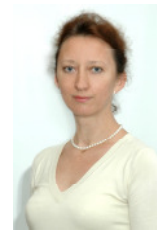